\documentstyle[12pt,epsf]{article}

\begin{document}

\newcommand{\sheptitle}
{On fixed points and fermion mass 
structure from large extra dimensions} 

\newcommand{\shepauthor}
{S. A. Abel$^a$ and S. F. King$^b$}

\newcommand{\shepaddress}
{
$a$ Theory Division, CERN, CH-1211 Geneva 23, Switzerland\\
$b$ Dept of Physics and Astronomy,
University of Southampton, SO17 1BJ, UK}

\newcommand{\shepabstract}
{We examine the fixed point behaviour of Yukawa couplings 
in supersymmetric theories with varying numbers of dimensions.
We show that Pendleton-Ross fixed point behaviour 
is greatly amplified in the MSSM with no extra dimensions and 4 
extra $5+\overline{5}$ multiplets or 
the MSSM with one extra large dimension and 3 extra $5+\overline{5}$ 
multiplets. We also show that power law running in models with 
large extra dimensions can give a hierarchical set of quasi-fixed points 
for the Yukawa couplings in a manner which is similar to
the Froggatt-Nielsen mechanism. However, we also point out 
the limited perturbative domain in models with power law running.}

\begin{titlepage}
\begin{flushright}
CERN-TH/98-311\\
hep-ph/9809467\\
\end{flushright}
\vspace{0.5in}
\begin{center}
{\large{\bf \sheptitle}}
\bigskip \\ \shepauthor \\ \mbox{} \\ {\it \shepaddress} \\ 
\vspace{0.5in}
{\bf Abstract} \bigskip \end{center} \setcounter{page}{0}
\shepabstract
\vspace{0.5in}
\begin{flushleft}
CERN-TH/98-311\\
\today
\end{flushleft}
\end{titlepage}

\def\sspace{\baselineskip = .16in}
\def\dspace{\baselineskip = .30in}
\def\beq{\begin{equation}}
\def\eeq{\end{equation}}
\def\bea{\begin{eqnarray}}
\def\eea{\end{eqnarray}}
\def\bq{\begin{quote}}
\def\eq{\end{quote}}
\def\ra{\rightarrow}
\def\lra{\leftrightarrow}
\def\ups{\upsilon}
\def\bq{\begin{quote}}
\def\eq{\end{quote}}
\def\ra{\rightarrow}
\def\un{\underline}
\def\ov{\overline}

\newcommand{\plb}[3]{{{\it Phys.~Lett.}~{\bf B#1} (#3) #2}}
\newcommand{\npb}[3]{{{\it Nucl.~Phys.}~{\bf B#1} (#3) #2}}
\newcommand{\prd}[3]{{{\it Phys.~Rev.}~{\bf D#1} (#3) #2}}
\newcommand{\ptp}[3]{{{\it Prog.~Theor.~Phys.}~{\bf #1} (#3) #2}}
\newcommand{\hepph}[1]{{\tt hep-ph/#1}}
\newcommand{\hepth}[1]{{\tt hep-th/#1}}
\newcommand{\leqsim}{\,\raisebox{-0.6ex}{$\buildrel < \over \sim$}\,}
\newcommand{\geqsim}{\,\raisebox{-0.6ex}{$\buildrel > \over \sim$}\,}
\newcommand{\be}{\begin{equation}}
\newcommand{\ee}{\end{equation}}
\newcommand{\ba}{\begin{eqnarray}}
\newcommand{\ea}{\end{eqnarray}}
\newcommand{\nn}{\nonumber}
\newcommand{\ie}{\mbox{{\em i.e.~}}}
\newcommand{\mpl}{\mbox{$M_{pl}$}}
\def\gev{\,{\rm GeV}}
\def\dd{\mbox{d}}
\def\etal{\mbox{\it et al }}

\section{Introduction}

One interesting prospect for physics beyond the standard 
model is that at high energies, beyond the reach of 
current experiment, hidden extra dimensions open up, revealing 
themselves through, for example, the appearance of 
Kaluza-Klein modes \cite{ant1,tony,shiu}.
Above the first Kaluza-Klein threshold there would be  
enhancement in the effects of the renormalisation
group. For example, in Ref.\cite{tony} it was recently 
noted that consistent unification can occur 
at much lower scales because, by changing the classical dimension of the 
gauge couplings, large extra dimensions cause the (effective)
renormalisation group equations to run as a power law with energy scale. 

In this paper we follow up a different 
point which was emphasised in Refs.\cite{graham,tony}; 
that power law running may lie behind
the extraordinary hierarchies observed in the fermion mass matrices. 
Ref.\cite{tony} presented some interesting 
proposals in which large extra dimensions
may provide an explanation for these hierarchies and hence an 
alternative to the Froggatt-Nielsen mechanism.
It was noted, for example, that when the effective (one loop) renormalisation 
group equations (RGE's) are run upwards to the unification scale, 
the Yukawa couplings appear to have a common Landau pole.
Hence it was suggested that the Landau pole could be 
indicative of `Yukawa unification'. 

Here we shall carry out a more detailed analysis of 
these suggestions concentrating in particular on their
domain of (perturbative) validity. Generally we shall find that, 
if the Yukawa hierarchies are generated from scratch by the renormalisation 
group, then perturbation theory breaks down well below the unification scale.
However we shall present a case in which there is a hierarchical 
set of fixed points which {\em can} be calculated within perturbation 
theory. 

Our approach is to consider the existence of quasi-fixed 
points (QFP's) and their domain of attraction. This is because, in any 
scenario of Yukawa `unification'\footnote{We use this term loosely to 
mean a scenario in which the Yukawa hiererachy is generated by the 
renormalisation group running alone}, at least some of the 
Yukawa couplings must 
have common Landau poles and hence, by definition, will be close to QFP's. 
First, in the following section, we revisit the (effective) 
RGE's for the gauge and Yukawa couplings for cases where 
different generations `feel' different numbers of large compact 
extra dimensions via Kaluza-Klein (KK) modes. 
At first sight, it does indeed appear that some rather subtle features
of the renormalisation group in the MSSM (such as fixed point
behaviour) could be made dominant by power law running. 

In section 3 we make a general examination of the 
expected fixed point behaviour (both Pendleton-Ross and quasi-fixed)
when we extend the MSSM by allowing different particles to feel different 
numbers of extra dimensions. 

For the Pendleton-Ross fixed points we find that when both gauge and 
Yukawa couplings run as a power law (even if the powers are 
different), attraction to these fixed points 
is generally neither more nor less marked than in 
the MSSM because the lower unification scale exactly compensates for the 
enhanced running. 
In fact the parameters describing the properties of the 
Pendleton-Ross fixed point (the domain of attraction for example) 
can be expressed 
in terms of the running gauge couplings and so ultimately depend 
only on the strength of these couplings at the unification scale. 
Moreover, perturbativity limits place strong constraints on the 
gauge couplings, so that the allowed fixed point behaviour increases 
as the number of extra dimensions goes {\em down}. 
In section 3 we also highlight a case where the fixed point behaviour is 
stronger than usual; when the beta function for the strong 
coupling is $b_3=1 $ 
the domain of attraction and focussing to fixed points is greatly 
increased. This corresponds to 
the MSSM with one extra large dimension and 3 extra $5+\overline{5}$ 
multiplets, or the MSSM with no extra dimensions and 4 
extra $5+\overline{5}$ multiplets. In both these cases the fixed point 
behaviour can be very strong even for quite modest gauge couplings 
at the GUT scale. However we find that, if perturbativity limits
are satisfied, the maximum hierarchy which could be generated by 
Pendleton-Ross fixed points is $\sim 30$.

We also examine the running to QFP's. This is the type of 
behaviour which was indicated in Ref.\cite{tony} where an emphasis 
was placed on the existence of Landau poles. For the case where we 
just extend the MSSM by allowing extra dimensions, we find that
running to fixed points cannot generate a significant hierarchy 
within the perturbative regime. 
However we also present a model with additional singlets, in
which the results are more promising. This model does have 
Standard-Model-like hierarchies in
the QFP's which can be calculated within perturbation theory. 
We add, however, the caveat that although the
hierarchical QFP's lie within the perturbative regime they should
be thought of as boundary conditions to the perturbative theory.
It is (currently) not possible to calculate the domain of attraction 
of these fixed points. 

\section{RGE's with extra dimensions.}

We first write down the MSSM RGE's in a useful matrix form which
we will shortly generalise;
\ba
4\pi \frac{\dd h_U}{\dd t} &=& -h_U.{N}_U
-{N}_Q.h_U -({N}_{H_2})h_U 
\nn\\
4\pi \frac{\dd h_D}{\dd t} &=& -h_D.{N}_D
-{N}_Q.h_D -({N}_{H_1})h_D 
\nn\\
4\pi \frac{\dd h_E}{\dd t} &=& -h_E.{N}_E
-{N}_L.h_E -({N}_{H_1})h_E 
\ea 
where  
\ba
N_{H_1}&=& 
\left( \frac{3}{2}g_2^2 +\frac{3}{10}g_1^2 
\right) -3Tr(h_D^\dagger h_D)-Tr(h_E^\dag h_E) \nn \\
N_{H_2}&=& 
\left( \frac{3}{2}g_2^2 +\frac{3}{10}g_1^2 
\right) -3Tr(h_U^\dagger h_U) \nn \\
N_{Q}&=& 
\left(\frac{8}{3}g_3^2 + \frac{3}{2}g_2^2 +\frac{1}{30}g_1^2 
\right)
-h_Uh_U^\dagger-h_Dh_D^\dagger\nn\\
N_{U}&=& 
\left(\frac{8}{3}g_3^2 +\frac{8}{15}g_1^2 
\right)
-2 h_U^\dagger h_U 
\nn\\
N_{D}&=&
\left(\frac{8}{3}g_3^2 +\frac{2}{15}g_1^2 
\right)
-2 h_D^\dagger h_D
\nn\\
N_{L}&=&
\left( \frac{3}{2}g_2^2 +\frac{3}{10}g_1^2 
\right)
-h_Eh_E^\dagger
\nn\\
N_{E}&=&
\left(\frac{6}{5}g_1^2 
\right)
-2 h_E^\dagger h_E. 
\ea
In these equations the parameters run logarithmically with 
scale; 
\be 
t=\frac{1}{4\pi} \log \frac{\Lambda}{\mu}.
\ee 
The $N_F$ factors come from the wavefunction renormalisation
of the $F$ superfield corresponding to one-loop diagrams with 
either matter, or matter-plus-gauge particles in the loop.

In the rest of this paper we shall closely follow the models 
described in Ref.\cite{tony}.
We shall assume that the massless fields appear only as $N=1$ 
supermultiplets. If they have a KK tower of $N=2$ states this is because 
half of the states lack a zero mode as they are odd under a $Z_2$ 
orbifolding. If they do not have a KK tower of states this is because they 
live, for example, at the fixed points of the orbifold or 
at the intersection of two branes (for more details see Ref.\cite{tony}).
The $N=2$ hypermultiplets contain the usual $N=1$ part plus mirror 
partners with the opposite charges (so that, for example, we can write 
mass terms for them). The $N=2$ vectormultiplets consists of the $N=1$ 
vectormultiplet plus  an additional $N=1$ chiral multiplet which 
corresponds to the longitudinal degree of freedom of the massive
gauge bosons.  
Yukawa couplings involving the hypermultiplets 
are restricted in $N=2$ models in order to preserve 
the $N=2$ supersymmetry. However we will for the moment 
assume that a combination 
of trilinear $N=2$ couplings and higher order interactions between $N=2$ 
hypermultiplets generates a set of Yukawa couplings at the GUT scale. 

Now let us consider the contribution to beta functions 
coming from extra sets of $N=2$ hypermultiplets appearing 
in towers of KK modes. This is meaningful if the full 
(but non-renormalisable) higher dimensional theory is well approximated by 
a truncated model Ref.\cite{tony}. (One should bear in mind 
that the `RGE's' we will derive do not have the same physical interpretation 
as in renormalisable models, in that they express the dependence of the 
physical parameters on the cutoff.) For example one has to
assume that higher dimensional operators do not play a significant role 
in the evolution of the gauge and Yukawa couplings\footnote{We would 
like to thanks R.Rattazzi for conversations on these points.}.
 
First the diagrams with matter only. These diagrams can have either one or two 
internal KK-modes. If there is only one, then, each time we pass a 
KK threshold, the diagram contributes the same as the equivalent 
MSSM diagram regardless of whether the external state has, or does not have, 
KK modes, and independently of the KK number of the internal mode. This is 
because the modes which live at orbifold fixed points (or at the intersection 
of two branes) do not conserve KK number since translational invariance 
in the extra dimensions is broken. These states, since they 
are located at fixed points, are eigenstates of position and therefore 
couple with the same coefficient to all the momentum eigenstates.
(Momentum is not really broken in these interactions 
but is absorbed by the unknown dynamics which maintains the brane or 
orbifold configuration.)
If there are two internal KK-modes and the external mode is the zero mode
of a tower of KK states, then the diagram also contributes the same as 
the equivalent MSSM diagram since now KK number must be conserved at the 
vertex. If, however, there are two internal KK modes and the external state 
lives at a fixed point, then every time we pass a KK threshold 
we must sum over all combinations of the two internal KK modes 
corresponding to that threshold. 

Now the diagrams with matter-plus-gauge in the loop. 
When we cross a KK threshold, they 
give the same contribution as the equivalent MSSM diagram 
if the matter multiplet appearing in the loop has no KK tower of states. 
This is 
because of a cancellation between diagrams involving only $N=2$ multiplets
(the same cancellation that gives vanishing wavefunction renormalisation 
for hypermultiplets in unbroken $N=2$ theories).

In general therefore one expects two types of contribution to 
the beta functions at a given threshold: those which involve a
single summation and and those which involve a double summation
over KK modes below that threshold. We shall denote these with a single 
and double tilde respectively and shall, for the moment, neglect
the usual MSSM contributions. 
It is necessary to separate these two contributions because as we shall 
see they give different scale dependence in the RGE's. In 
an effective theory with truncated KK modes the general form of the 
RGE's can be expressed as 
\be 
4\pi \frac{\dd x_i}{\dd t }=x_i\beta_{ij}x_j
\ee
where $x_i$ contains the whole set of $h_i^2$ or $g^2$.
(Actually they can only be written like this when the Yukawa couplings 
are diagonal, but the discussion is the same for non-diagonal Yukawas).
The beta functions are now rapidly changing with scale and to estimate 
this we count the number of KK mode contributions below a 
cut-off $\Lambda$. If each KK mode is separated by 
a scale $\mu_0$ and each adds $\tilde{b}_{ij}$ or $\tilde{\tilde{b}}_{ij}$
to the beta functions, then the beta function for a given $\Lambda $ is 
approximately 
\be 
\beta_{ij}(\Lambda)= 
X_{\delta}\left(\frac{\Lambda}{\mu_0}\right)^\delta\tilde{b}_{ij}
+X_{\delta}^2\frac{(\delta !)^2}{(2\delta ) !}
\left(\frac{\Lambda}{\mu_0}\right)^{2\delta}
\tilde{\tilde{b}}_{ij}
\ee
where $\delta $ is the number of large extra dimensions,
$X_\delta $ is volume of the unit sphere in $\delta$ dimensions,
\be 
X_\delta =\frac{\pi^{\delta/2}}{\Gamma(1+\delta/2)},
\ee
and where we neglect the contribution from the usual MSSM states. 
(We are, for the moment, assuming that all particles feel the 
same number of extra dimensions.
When the gauge and Yukawa couplings feel different numbers of 
dimensions, we shall take $t_\delta $ to be the number of dimensions
felt by the {\em gauge} couplings.)  
The infinitessimal solutions to the RGE's at a scale $\Lambda $ 
are then given by 
\ba
4\pi \delta x_i &=& \frac{1}{4\pi}\frac{\delta\Lambda}{\Lambda}
 X_\delta 
\left(\frac{\Lambda}{\mu_0}\right)^\delta 
x_i\tilde{b}_{ij}x_j
+
\frac{1}{4\pi}\frac{\delta\Lambda}{\Lambda}
 X^2_\delta \frac{(\delta !)^2}{(2\delta ) !}
\left(\frac{\Lambda}{\mu_0}\right)^{2\delta} 
x_i\tilde{\tilde{b}}_{ij}x_j\nn\\
&=&\delta t_\delta \left(
x_i\tilde{b}_{ij}x_j +
4\pi \delta \frac{(\delta !)^2}{(2\delta ) !}
t_\delta x_i\tilde{\tilde{b}}_{ij}x_j\right)
\ea
where we have defined 
\be
\label{tdelta}
t_\delta = \frac{X_\delta}{4\pi\delta } \left(\frac{\Lambda}
{\mu_0}\right)^\delta.
\ee
Note that when $\delta =0 $, $t_\delta$ assumes the usual logarithmic form.
In the limit that $\mu_0 \ll \Lambda $ we can approximate the evolution 
with a power law running set of RGE's.

Gathering together all of these results, we can now write down 
the combined effect of the KK thresholds. We shall assume that the 
higgs fields have towers of KK states. In addition, 
to treat models with
different numbers of generations having KK towers, we define
\be
\Omega =\left(\begin{array}{ccc}
a_1&0&0\nn\\
0&a_2&0\nn\\
0&0&a_3
\end{array}\right),
\ee
where $a_i=0(1)$ when generation $i$ is(is not)
the zero mode of a tower of KK states. 
The contributions to the field renormalisation 
from the single summation diagrams are
\ba
\label{rges0}
\tilde{N}_{H_1}&=& -3 (Tr(h_Dh_D^\dagger) - Tr(h_D\Omega h_D^\dagger\Omega ))
-  (Tr(h_Eh_E^\dagger) - Tr(h_E\Omega h_E^\dagger\Omega ))\nn\\
\tilde{N}_{H_2}&=& -3 (Tr(h_Uh_U^\dagger) - 
Tr(h_U\Omega h_U^\dagger\Omega ))\nn\\
\tilde{N}_{Q_{ij}}&=& \Omega_{ij}
\left(\frac{8}{3}g_3^2 + \frac{3}{2}g_2^2 +\frac{1}{30}g_1^2 
\right)
-h_Uh_U^\dagger-h_Dh_D^\dagger\nn\\
&& + \Omega h_U (1-\Omega) h_U^\dagger \Omega +
 \Omega h_D (1-\Omega) h_D^\dagger \Omega 
\nn\\
\tilde{N}_{U_{ij}}&=& \Omega_{ij}
\left(\frac{8}{3}g_3^2 +\frac{8}{15}g_1^2 
\right)
-2 h_U^\dagger h_U + 2 \Omega h_U^\dagger (1-\Omega) h_U \Omega 
\nn\\
\tilde{N}_{D_{ij}}&=& \Omega_{ij}
\left(\frac{8}{3}g_3^2 +\frac{2}{15}g_1^2 
\right)
-2 h_D^\dagger h_D+ 2 \Omega h_D^\dagger (1-\Omega) h_D \Omega 
\nn\\
\tilde{N}_{L_{ij}}&=& \Omega_{ij}
\left( \frac{3}{2}g_2^2 +\frac{3}{10}g_1^2 
\right)
-h_Eh_E^\dagger+  \Omega h_E (1-\Omega) h_E^\dagger \Omega 
\nn\\
\tilde{N}_{E_{ij}}&=& \Omega_{ij}
\left(\frac{6}{5}g_1^2 
\right)
-2 h_E^\dagger h_E + 2 \Omega h_E^\dagger (1-\Omega) h_E \Omega ,
\ea
and the double summation contributions are
\ba
\tilde{\tilde{N}}_{Q_{ij}}&=& - \Omega h_U (1-\Omega) h_U^\dagger \Omega 
- \Omega h_D (1-\Omega) h_D^\dagger \Omega 
\nn\\
\tilde{\tilde{N}}_{U_{ij}}&=& - 2 \Omega h_U^\dagger (1-\Omega) h_U \Omega 
\nn\\
\tilde{\tilde{N}}_{D_{ij}}&=& - 2 \Omega h_D^\dagger (1-\Omega) h_D \Omega 
\nn\\
\tilde{\tilde{N}}_{L_{ij}}&=& -  \Omega h_E (1-\Omega) h_E^\dagger \Omega 
\nn\\
\tilde{\tilde{N}}_{E_{ij}}&=& - 2 \Omega h_E^\dagger (1-\Omega) h_E \Omega .
\label{rges1}
\ea
The RGE's can then be written, 
\ba
4\pi \frac{\dd h_U}{\dd t_\delta} &=& -h_U.(\tilde{N}_U+
4\pi \delta \frac{(\delta !)^2}{(2\delta ) !}
 t_\delta 
\tilde{\tilde{N}}_U)-
(\tilde{N}_Q+4\pi \delta \frac{(\delta !)^2}{(2\delta ) !}t_\delta
 \tilde{\tilde{N}}_Q).h_U\nn\\
4\pi \frac{\dd h_D}{\dd t_\delta} &=& -h_D.(\tilde{N}_D+
4\pi \delta \frac{(\delta !)^2}{(2\delta ) !}
t_\delta 
\tilde{\tilde{N}}_D)-
(\tilde{N}_Q+4\pi \delta \frac{(\delta !)^2}{(2\delta ) !}
 t_\delta \tilde{\tilde{N}}_Q).h_D\nn\\
4\pi \frac{\dd h_E}{\dd t_\delta} &=& -h_E.(\tilde{N}_E+
4\pi \delta \frac{(\delta !)^2}{(2\delta ) !}
t_\delta 
\tilde{\tilde{N}}_E)-
(\tilde{N}_L+4\pi \delta \frac{(\delta !)^2}{(2\delta ) !}
 t_\delta \tilde{\tilde{N}}_L).h_E\nn\\
4\pi \frac{\dd g_A}{\dd t_\delta} &=& \tilde{b}_A g_A^3 ,
\label{RGEs2}
\ea
where~\cite{tony} 
\be
\tilde{b}_A=\left(\frac{3}{5},-3,-6\right)+
(4 \eta + n_{5+\overline{5}})\left(1,1,1\right).
\ee
In the above, $\eta = 3-a_1-a_2-a_3$ counts the number of generations 
with KK modes and we have allowed for the possibility of 
$n_{5+\overline{5}}$ sets of $5+\overline{5}$ multiplets. 

These RGE's describe the integrated effect of 
the KK thresholds. In what follows we will assume that they can be resummed 
in the usual way to get a better approximation than the leading log 
approximation used in Ref.\cite{tony}. In doing this we note that the 
conventional resummation of one loop Yukawa RGE's can be understood as 
a series of nested and one-particle-reducible field renormalisation 
diagrams. Inspecting 
these diagrams shows that these operators and only these appear 
at higher order. In particular (at this level) we never need to know 
about the Yukawa couplings of the mirror partners of the usual MSSM 
fields which appear in the hypermultiplets.
We stress that `running the RGE's' has a different physical 
interpretation to the usual procedure in renormalisable field theory
although the mathematical procedure is the same. 
Here the RGE's represent a summation of diagrams which give 
{\em finite} corrections corresponding to KK contributions below the
cut-off $\Lambda $. 

We are also assuming that the perturbation theory is still valid 
over the region of integration; this point will be discussed in the 
following section.
Finally we must assume that the effect of these RGE's is almost 
continuous or in other words that there are many KK modes before unification.
(In Ref.\cite{tony} this last approximation was shown to be valid  
since $t_\delta \sim 20 $.) 

\section{Fixed points with extra dimensions; the plot thins.}

Before considering the full running of the flavour dependent RGE's we first 
discuss general renormalisation group behaviour with large extra dimensions.
In the previous section we saw that the parameters scale as a power law 
with energy scale as opposed to the familiar logarithmic running seen in 
the MSSM. In addition, when particles feel different numbers of dimensions
the power is different. This naturally leads one to suppose that scaling 
effects 
will be very strong, and that they could {\em by themselves} be responsible 
for the hierarchies observed in the fermion mass matrices~\cite{graham,tony}. 

When renormalisation group effects are strong, the main features 
of the running are determined, more or less, by the presence of fixed points.
There are two kinds of fixed points which are familiar from the usual MSSM; 
they are Pendleton-Ross fixed points (PRFP's) 
and quasi-fixed points (QFP's)\cite{fps,fps3,fps2}.
The PRFP is the true fixed point in the sense that couplings are 
attracted towards it in the Infra-Red. However, in the MSSM this is 
not the dominant feature. In the MSSM QFP's are the dominant feature 
(for the top-quark Yukawa) because the top mass is relatively close to the 
perturbativity limit, and the QFP corresponds to the value of the Yukawa 
coupling when there is a Landau pole at the GUT scale. 

In this section we shall give a general discussion of what form these two 
types of behaviour take when there are large extra dimensions
using a simple example which will allow us to deduce all the fixed 
point behaviour in the more complicated cases of interest.
We shall find,
rather surprisingly, that PRFP behaviour is, in general, not expected to be a 
significant factor. Of course, when compared to the energy scale, 
the running to the PRFP is indeed very strong since in $4+\delta $ dimensions 
the Yukawa couplings have classical dimension. 
So, however, do the gauge couplings
(unless the they feel none of the extra dimensions or the contributions 
to their 
beta functions vanish) and it is these that set the scale of unification.
It is not likely therefore, that PRFP's play a large part in the generation 
of fermion hierarchies. We also find that QFP's are 
not expected to be significant if the only extension we make to the MSSM 
is to allow some particles to feel extra dimensions. A number of 
other scenarios can have hierarchical QFP's however, and we will examine 
one case which involves extra singlets. 

\subsection{PRFP's in a generic case}

To be more specific, let us examine the renormalisation of a Yukawa
coupling, $h_t$ (which, for the sake of argument, we shall call the top 
quark Yukawa), whose RGE is given by 
\be
\label{generic}
4\pi \frac{\dd h_t}{\dd t_\delta} = h_t\left(
a h_t^2 -c_A g_A^2
\right).
\ee
If we put $a=3$ and $c_A=(17/30,3/2,16/3)$ then this is the 
$h_{U_{33}}$ RGE of the previous section when $h_{U_{33}}$ is dominant. 
This equation is also of the same form as that in the usual MSSM ($\delta=0$) 
except in that case $a=6$ and $c_A=(13/15,3,16/3)$, and so we can use the 
same solutions. Defining 
\ba
r_A&=&
\alpha_0/ \alpha_A \nn\\
R_t &=& 
\frac{h_t^2 }{g_3^2},
\ea
where we use subscript-0 to denote values at the unification 
scale, we find the solutions 
\ba 
\label{Rsoln1}
r_A&=& 1-2 \tilde{b}_A \alpha_0 \Delta t_\delta \nn\\
\frac{1}{R_t}&=&\frac{1}{r_3 R_0\Pi  } - J
\ea
where 
\ba 
\Pi (r_A) &=& \prod_A r_A^{c_A/\tilde{b}_A} \nn\\
J &=&\frac{a}{\tilde{b}_3 r_3\Pi}\int^1_{r_3}  \Pi \dd r_3'  .
\ea
To discuss the fixed point behaviour it helps to define an 
`instantaneous' fixed point, $R^*_t(r)$, which can be thought of as 
the value of $R_t$ which is approached when the gauge couplings are 
renormalising very slowly. That is 
\be 
\label{fixedp}
\frac{1}{R^*_t}=\frac{1}{r_3 R_t^*\Pi  } - J
\ee
or 
\be 
\label{r*}
R^*_t=\frac{1}{J} \left( \frac{1}{r_3 \Pi} -1 \right).
\ee
Substituting into $R_t$ gives 
\be 
\label{Rsoln}
R_t=\frac{R^*_t }{1+\Delta \left( \frac{R^*_t}{R_0}-1 \right)}.
\ee
where 
\be 
\label{delta}
\Delta (r_3) = \frac{1}{r_3\Pi}
\ee
defines the domain of attraction of the fixed point. In the usual MSSM 
$\Delta \approx r_3^{7/9}\approx 0.37^{7/9} \approx 1/2 $.

The virtue of $R^*_t(r_3)$ is that it separates the minor effects of 
gauge renormalisation from the `PRFP behaviour' which is 
given by $\Delta $. (We emphasise that this is not a new kind of fixed point. 
For example, if the gauge coupling renormalisation is as fast as that of the 
Yukawas then, even if $R_t$ is set to be $R^*_t(r_3)$ at some scale, it will 
leave the $R_t^*(r_3) $ line.)

The PRFP corresponds to taking the infra-red limit 
(\ie $r_3\rightarrow \infty$ for positive $b_3$ or $r_3\rightarrow 0$ in 
the MSSM) to find $1/(r_3\Pi) \rightarrow 0$ to give, for example,
\be
R^*_t(0) = -1/J(0) = 7/18   
\ee
in the MSSM.
Here we can identify some interesting new cases where the domain of attraction 
is very large. For example, 
we can add $4\times (5+\overline{5})$ multiplets to the MSSM 
to make $\tilde{b}_3=+1$. In this case we find $R^*_t(0)=19/18$ and  
\be 
\label{19/3}
\Delta \approx r_3^{-19/3}.
\ee
Since positive $\tilde{b}_3$ means stronger coupling at unification,
$r_3$ can be $3$ for example. This gives a very large domain of
attraction since now $\Delta\approx 10^{-3}$ thanks to the large power
appearing above. In fact $\tilde{b}_3=+1$ appears to be the optimum
case. The top quark Yukawa is, in this instance, rapidly focussed
(from above {\em and below}) to the running value given by
$R^*_t(r_3)$ in Eq.(\ref{fixedp}) as shown in Fig.(1). (This behaviour
is unchanged by two loop corrections.)  In models with one large extra
dimension the equivalent would be to have one generation with KK-modes
and 3 additional $(5+\overline{5})$ multiplets. Then, for a dominant
third generation, we still have Eq.(\ref{19/3}) with $R^*_t(0)=19/9$,
and we conclude that in this case the extra dimensions have not increased 
the running to PRFP's.

\begin{figure}
\label{fig1}
\vspace*{-4in}
\hspace*{-2.0in}
\epsfysize=12in
\epsffile{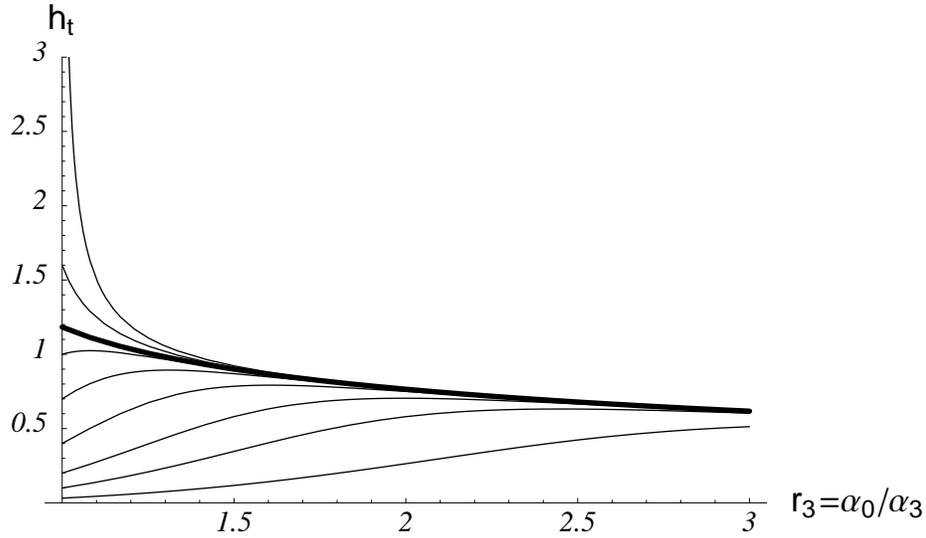}
\vspace{-2in}
\caption{Renormalisation of the top quark Yukawa in the MSSM with 
$4\times (5+\overline{5})$ multiplets. The bold line shows $R^*_t(r_3)$.
Its infra-red limit, $R^*_t(\infty)$, corresponds to the Pendleton-Ross 
fixed point. The highest line corresponds to the quasi-fixed point of Hill.
Two loop effects do not change the diagram significantly. The lowest 
line corresponds to a $h_t(M_{GUT})=0.03$.} 
\end{figure}

What about the cases where there are additional powers of $t_\delta $ in the 
RGE's or when, as in the RGE's of the previous section, there are double 
summations over KK modes? Even in these cases (even more counter-intuitively) 
the fixed point behaviour is not particularly enhanced. 
Indeed when we solve the general equation, 
\be
4\pi \frac{\dd h_t}{\dd t_\delta} = h_t\left(
a(t_\delta) h_t^2 -c_A g_A^2
\right),
\ee
where now $a(t_\delta )$ is any function of $t_\delta $, we find a
solution given by Eqs.(\ref{r*}),(\ref{Rsoln}),(\ref{delta}) with $J$
replaced by
\be 
\label{general}
J_a =\frac{1}{\tilde{b}_3 r_3 \Pi}\int^1_{r_3}  a(t_\delta(r'_3))\Pi  
\dd r_3'  
\ee
where 
\be 
\label{teqn}
t_\delta (r_3) = t_\delta (M_{GUT}) + \frac{1-r_3}{2\tilde{b}_3 \alpha _0 }.
\ee
Hence, only $R^*_t(r_3)$ is changed by any extra powers of $t_\delta $ and 
the domain of attraction, $\Delta (r_3)$, remains the 
same. Moreover it is not possible to consistently have a hierarchical set of 
PRFP's within perturbation theory.

Let us demonstrate the implications for a simple example.  For the
terms coming from double summations over KK modes in the previous
section, there is a doubled power of $t_\delta $.  We plot the
resulting renormalisation (with $a(t_\delta )=6 t_\delta $) in Fig.(2)
for the MSSM with a single extra dimension and 3 additional
$(5+\overline{5})$ multiplets. The extra power has not increased the
fixed point behaviour significantly. (We stop the renormalisation at
low scale at the mass of the lightest KK mode ($\Lambda=\mu_0$).)

\begin{figure}
\label{fig2}
\vspace*{-4in}
\hspace*{-2.0in}
\epsfysize=12in
\epsffile{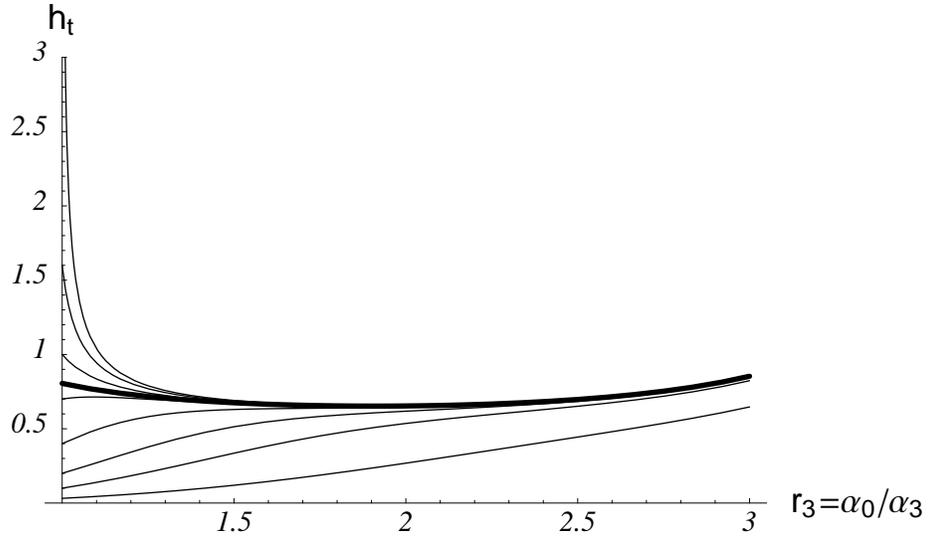}
\vspace{-2in}
\caption{Renormalisation of a Yukawa coupling 
with double power law scaling with one extra dimension 
and $3\times (5+\overline{5})$ multiplets. 
We choose $ 2\alpha _0 \tilde{b}_3 = 1 $ and $t_\delta (M_{GUT})= 2 
+ 1/2\pi $ (so that $r_3=3$ corresponds to $\Lambda = \mu_0 $).} 
\end{figure}

From the above discussion we conclude that large extra dimensions do
{\em not} generically lead to enhanced PRFP behaviour.  The power law
running to PRFP's is accompanied by an equally rapid running of the
gauge couplings so that the net result is merely that unification (or
loss of perturbativity) is reached at a much lower energy scale. We
also see that PRFP behaviour is a dominant feature of {\em strong}
unification (larger $r_3$ gives stronger 
attraction)~\cite{graham,strong1,strong2}.  In models with
extra dimensions the PRFP behaviour is therefore ultimately limited by
perturbativity constraints. Generally we require that $n+1$-loop diagrams 
contribute less than $n$-loop diagrams, which gives~\cite{strong2} 
\be 
N \tilde{b}_3\alpha_0 \leqsim 4\pi ,
\ee
where $N$ is the number of degrees of freedom contributing to the 
$\beta$-function. In this case 
\be
N=X_\delta \left( \frac{\Lambda}{\mu_0} \right) ^\delta. 
\ee
Perturbativity then requires 
\be 
\tilde{b}_3 \delta \alpha_0 \left|
\Delta t_\delta \right| \leqsim 1
\ee
and hence 
\be
r_3=1-2\tilde{b}_3 \alpha_0 \Delta t_\delta \leqsim 1+2/\delta.
\ee
So $r_3\approx 3 $ is the perturbativity limit for 
$\delta > 0$. On these quite general grounds therefore, it seems unlikely 
that one could generate a hierarchy of greater than $3^{19/6}\approx 30 $ 
from the effects of PRFP's alone. 

\subsection{QFP induced hierarchies}

Now we turn to QFP behaviour which happens in an entirely different region of 
parameter space. Here we shall find that hierarchical QFP's can exist within 
perturbation theory. The QFP corresponds to taking $R_0\rightarrow \infty $ 
in Eq.(\ref{Rsoln}) and hence, 
\be 
R^{QFP}_t\approx \frac{R^*_t(0) }{1-\Delta(r_3)}.
\ee
Now consider, simply as an example, what happens when the gauge
couplings feel none of the extra dimensions but the Yukawa coupling
does?  This could occur if there were extra multiplets in the KK
levels so that the contributions to the gauge beta functions come from
complete $N=4$ multiplets and hence cancel, but the Yukawa couplings
still receive contributions from towers of $N=2$ multiplets. (Since
this is just an example we shall not go to the trouble of actually
building such a model.)  Here one would expect power law running to
have a significant impact and indeed it does. In this case we can work
with the usual logarithmic variable, $t_\delta\equiv t$, and put
\be 
a(t(r_3)) \approx  X_\delta \left(\frac{M_{GUT}}{\mu_0} \right)^ \delta 
e^{16 \pi \delta (r_3 -1)},
\ee 
where we have taken $\alpha_0=1/24$. The prefactor means that 
\be 
J_a \approx -\frac{3}{16 \pi \delta } \left(
\frac{M_{GUT}}{\mu _0} \right) ^\delta \Delta (r_3)
\ee
is potentially huge. Temporarily ignoring the question of perturbativity, 
Eq.(\ref{Rsoln1}) tells us that 
\be 
R_t^{-1}  = 
\Delta(r_3)\left(
 R_0^{-1} +\frac{3}{16 \pi \delta } \left(
\frac{M_{GUT}}{\mu _0} \right) ^\delta\right).
\ee
All of the the solutions are `pinched' to very small values by the QFP
solution near the GUT scale. Deviation from the QFP is usually
expressed with the parameter $\rho = R_t / R_t^{QFP}$. Here we find
\be 
\rho \approx 1- 
\frac{16 \pi \delta }{3 R_0}  \left(
\frac{M_{GUT}}{\mu _0} \right) ^{-\delta}.
\ee
So that, unless $R_0$ is extremely small, the Yukawa 
couplings effectively follow the {\em quasi}-fixed point solution.

We can see that QFP's will be important whenever the factor, $a(t_\delta)$ 
is large during the renormalisation. For power law running we might hope to be 
able to generate hierarchies if, for example, $a(t_\delta) =  t_\delta^n$; 
by examining $J_a$ above we find
\be 
\label{qfpvalues}
h_t = g_3 \sqrt{R_t} \sim
1/\sqrt{t^n_{\delta} (M_{GUT})}
\ee
In Fig.(3) we show the running for the Yukawa couplings when 
$\delta=1 $, $t_\delta =10 $, $a(t_\delta)=6 t^3_\delta $, and with the 
same values of $c_A$ as in Fig.(2).
The GUT scale values for the 
Yukawa couplings are the same as in Figs.(1,2), so in 
this case it appears that hierarchies of $10^2 $ in the Yukawa couplings 
have been generated. Note that when the gauge couplings are running 
as a power law we must respect the perturbativity constraints so that
QFP's are generally more important when $\alpha _0 $ is small and  
we have {\em weak} unification. 
 
Perturbativity constraints also apply to the Yukawa couplings, 
so that it might appear 
that there is some contradiction in having such large coefficients 
for the Yukawa couplings in the RGE's. As for the gauge couplings we should 
require that 
\be
t_\delta a(t_\delta) h_t^2 \leqsim 4 \pi .
\ee
However the rough QFP values in Eq.(\ref{qfpvalues}) satisfy this bound
if $t_\delta \leqsim 4\pi $.
Hence the QFP solutions should be thought of as boundary conditions.
Once the model has dropped into the perturbative regime the Yukawa 
couplings will run to these fixed points, however we cannot explain why 
the model drops into the perturbative regime in the first place (although 
we shall make some additional comments about this in the discussion). 

\begin{figure}
\label{fig3}
\vspace*{-4in}
\hspace*{-2.0in}
\epsfysize=12in
\epsffile{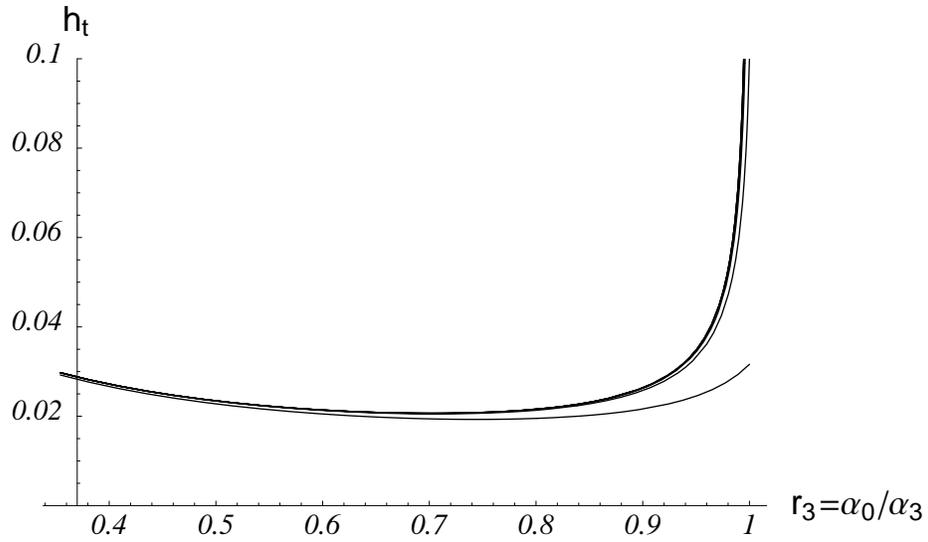}
\vspace{-2in}
\caption{Renormalisation of a Yukawa coupling 
which feels 4 extra dimensions of $N=2$ multiplets when the gauge couplings 
feel only one
extra dimension. We choose $\alpha_0=1/24$ and the same starting values 
for the Yukawa coupling as in Figs.(1,2). } 
\end{figure}

Now let us consider what models may exhibit such large 
$a(t_\delta )$. The RGE's we derived in the previous section were 
for cases in which
all particles feel the same number of extra dimensions. 
Here the largest terms are the double summation terms with 
$a(t_\delta)\sim t_\delta  $.  
Given the values of $t_\delta(M_{GUT})$ ($\leqsim 20 $) 
found in Ref.\cite{tony}, it is not possible 
to generate the required hierarchies unless the Yukawa couplings somehow 
feel more dimensions. This seems to be an unavoidable condition 
for generating large hiererachies from power law running. 

It does not seem possible to achieve 
this simply by having the particles feel different numbers of 
dimensions (at least within perturbation theory). For example, one might 
try having the gauge particles feel $\delta $ dimensions
and the matter and higgs particles feel $\delta'>\delta$ dimensions.
In this case we modify the Yukawa RGE's by multiplying all of the Yukawa terms 
in Eqs.(\ref{rges0},\ref{rges1}) by 
\be 
\frac{\dd t_{\delta'}}{\dd t_\delta}=
\frac{t_{\delta'}\delta'}{t_\delta \delta}
\ee
where
\be
t_{\delta '}= \frac{X_{\delta'}}{4 \pi\delta '}\left( 
\frac{4\pi\delta t_\delta}{X_\delta}\right)^{\delta'/\delta},
\ee
and by replacing $\delta $ by $\delta'$ on the RHS of Eq.(\ref{RGEs2}),
and $t_\delta $ by $t_\delta'$. 
However the gauge beta functions recieve contributions from all 
$\delta'$ dimensions of matter and Higgs particles; hence we should also 
multiply $\eta $ by 
$\frac{ t_{\delta'}\delta'}{t_\delta\delta } $ and thus the gauge 
couplings become strong before unification. In this case our previous 
perturbativity limits on 
the gauge couplings restrict $t_{\delta'}\leqsim 1$ and prevent 
it from generating a significant hierarchy. 

There are three possible ways we can allow the beta functions of the 
Yukawa couplings to feel more dimensions of KK modes; 
let the gauge particles interact with additional (vector and chiral) 
multiplets so that they feel entire $N=4$ multiplets rather than 
$N=2$ hypermultiplets from at least some of the extra dimensions; introduce 
additional non-perturbative gauge couplings for some of the 
fermions;
modify the Yukawa couplings themselves 
by adding extra singlets which are fully dynamical fields above the 
scale $\mu_0$~\cite{tony}. We shall now examine the last idea
(although much of the discussion will apply to the other two). 

We begin with a model in which none of the matter multiplets 
feel extra dimensions (the $\eta=0$ scenario of Re.\cite{tony})
and the Higgs and gauge multiplets feel one ($\delta=1$).
We add two singlets, which we call $\Phi$ and  $\overline{\Phi}$,
so that the Yukawa couplings are of the form
\be 
W= h_{F_{ij}} \left(\frac{\Phi}{\mu_0} \right)^{n_{F_i}}
\left(\frac{\overline{\Phi}}{\mu_0} \right)^{n_{\overline{F}_j}} 
F_i {\overline{F}_j}
 H_{u,d}.
\ee
We shall choose the $n_F$'s and $n_{\overline{F}}$'s 
to be functions of the generation numbers of 
the multiplets $F$ to introduce the required flavour dependence in 
the evolution of the couplings. Above the scale $\mu_0$ the 
wavefunction renormalisation diagrams are replaced by diagrams with 
$n_F+n_{\overline{F}}$ extra loops of $\Phi$ and $\overline{\Phi}$
particles.

The wavefunction renormalisation receives contributions which scale as 
$(\Lambda /\mu_0)^{3(n_F+n_{\overline{F}})+1}$ \cite{tony}
so that the RGE's of Eq.(\ref{RGEs2}) are modified by multiplying the Yukawa 
terms by an additional factor 
\be 
t_\delta ^{3 (n_{F_i}+n_{\overline{F}_j})}
\ee
where now $n_{F_i}+n_{\overline{F}_j}$ counts the number of singlets appearing 
in the leading diagram. For example, let $n_{F_i}$ and $n_{\overline{F}_j}$ be 
simply $3-i$ and $3-j$. To modify the RGE's we define 
\be 
\eta _{i}= t_\delta ^{3(3-i)}.
\ee
In the wavefunction renormalisation diagrams for the matter fields 
we then replace 
\ba 
(h_U h_U^\dagger )_{ij} &\rightarrow & \delta_{ij}\eta_i 
\sum_{k} h_{U_{ik}}\eta_k (h_U^\dagger)_{ki}\nn\\
(h_U^\dagger h_U )_{ij} &\rightarrow & \delta_{ij}\eta_i 
\sum_{k}  (h_U^\dagger)_{ik}\eta_k  h_{U_{ki}}
\ea
and similar for the down and lepton fields. In addition the Higgs, 
$\Phi $ and $\overline{\Phi}$ renormalisations are suppressed by one 
power of $t_\delta $ (since the matter multiplets do not have 
a tower of KK excitations) so we can safely neglect them. 
We now resum the RGE's in the usual way (again bearing in mind that 
the resummation is merely a convenient way of including 
finite threshold effects). 

We can estimate the QFP's as follows. Assume that initially 
the $h_U$ Yukawa couplings are all roughly the same size
at the GUT scale and take $t_\delta (M_{GUT})$ to 
be large (as it has to be in order to generate a hierarchy 
at all). We then 
approximate the RGE's using $\eta _1\gg \eta_2\gg\eta_3$. The $h_{U_{11}}$
RGE takes the same form as Eq.(\ref{generic}) with 
\be 
a_{11}(t_\delta) \approx 3 \eta_1^2 .
\ee
Its low-scale quasi-fixed value is $h_{U_{11}}\sim 1/\eta_1 $. 
When $h_{U_{11}}$ 
is near its quasi-fixed value, the RGE for $h_{U_{i1}}$ takes the same 
form as Eq.(\ref{generic}) with 
\be 
a_{1i}(t_\delta) \approx 3 \eta_1 \eta_i  ,
\ee
so that its quasi-fixed value is $h_{U_{1i}}\sim 1/\sqrt{\eta_1\eta_i} $,
and so on. The final form for the quasi-fixed $h_U$ is found to be 
\be 
\label{structure}
h_{U_{ij}} \sim 1/\sqrt{\eta_i\eta_j}
 = 
\left(
\begin{array}{ccc}
\epsilon^4 & \epsilon^3 & \epsilon^2  \nn\\
\epsilon^3   & \epsilon^2 & \epsilon    \nn\\
\epsilon^2   & \epsilon & 1          
\end{array}
\right)
\ee
where 
\be 
\epsilon \sim t_\delta (M_{GUT}) ^{-3/2} .
\ee
The final hierarchy is 
\be 
(h_t,h_c,h_u)\sim (1,\epsilon^2 ,\epsilon^4)
\ee
and the CKM matrix is of the form 
\be 
\label{ckmmatrix}
K
 \sim  
\left(
\begin{array}{ccc}
1 & \epsilon & \epsilon^2  \nn\\
\epsilon   & 1 & \epsilon    \nn\\
\epsilon^2   & \epsilon & 1          
\end{array}
\right).
\ee
With this ansatz the QFP's 
assume a structure with hierarchies similar to those in the 
Standard Model if $\epsilon \sim 0.1 $ and clearly, in this case, 
the structure with singlets is similar to
that one would have with the usual Froggatt-Nielsen mechanism.
However this is also true for two examples we have not considered 
that do not have additional singlets. 
Note that from the mass hierarchies we generally expect to get an 
estimate of the scale ($\mu_0$) at which new KK states will appear;
since $\epsilon \sim 0.1 $ we find $t_\delta \sim 10^{2/3} \approx 5 $
and hence $M_{GUT}\sim 15 \mu_0$ for $\delta=1$. Consulting Ref.\cite{tony}
we see that this implies 
\be
\mu_0 \sim 10^7 \gev.
\ee

The perturbativity discussion of the preceding section can be carried over 
directly to the present case and now, unfortunately, we again find that 
the (low scale) QFP's are close to the naive perturbative limit as 
is also the case for the two examples we did not consider. Perturbation theory 
does not allow us to follow the renormalisation of the Yukawa couplings 
very far from their low scale values. 

\section{discussion}

In this study we discussed various aspects of fixed point behaviour 
in theories both with and without extra dimensions. 
We found that in models where the strong gauge beta function is $b_3=+1$
(the MSSM with no extra dimensions and 4 
extra $5+\overline{5}$ multiplets or 
the MSSM with one extra large dimension and 3 extra $5+\overline{5}$ 
multiplets) the effects of Pendleton-Ross fixed points are greatly enhanced. 
However we argued that the maximum hierarchies that can be generated from 
Pendleton-Ross fixed points within perturbation theory are $\sim 30$. 

We also examined the effect of successive 
KK thresholds on the running of Yukawa couplings to QFP's.
We find that adding extra dimensions can enhance the 
effects of these fixed points appreciably; 
QFP's can be a dominant feature of 
the one loop `running'. With very simple assumptions one can generate 
Standard-Model-like hierarchies in the QFP's of the 
Yukawa couplings. Recent work on general questions regarding fixed 
points in supersymmetric theories \cite{fps2} leads us to believe 
that a similar fixed point structure exists for the soft 
supersymmetry breaking terms.

However we also highlighted some difficulties with this picture. Most 
importantly, perturbation theory is only valid when the Yukawa couplings 
are already near their (low-scale) quasi-fixed values. 
Hence, the low scale QFP's are close to the perturbative limit and 
have, for example, the same status 
as gauge couplings in strong unification models~\cite{strong1,strong2}.  
(Indeed, from this analysis, it seems likely that any attempt to 
generate significant Yukawa 
hierarchies through the renormalisation group will lead to a break 
down in perturbation theory.) 

Beyond perturbation theory it is very difficult to say anything about the 
domain of attraction of these fixed points and hence it is not possible 
to determine to what extent hiererachical QFP's actually 
constitute a prediction.
Despite this we find hierarchical QFP's intriguing
and deserving of more study. A hint that further progress may be possible 
(which we shall not explore here) 
lies in the fact that, for the model we presented, the dominant 
pieces of the beta functions of the Yukawa 
couplings appear to have signs which alternate with order in $h^2$. 
This follows from the fact that, at any order, diagrams with Yukawa 
vertices dominate over those with gauge vertices, so that the beta function 
involves the same set of diagrams as the Wess-Zumino model. 
Sign alternation is characteristic of asymptotic series which are 
Pad\'e-Borel summable and indeed using this technique for the Wess-Zumino 
model shows that the domain of  
attraction of the QFP's are substantial~\cite{fps3}. 
Coupled with the properties of the underlying $N=2$ properties 
this may allow domains of attraction which are larger
than the naive perturbative limit would imply. 

\bigskip

\section{Acknowledgements}
We are extremely grateful to Keith Dienes, Emilian Dudas, Tony Gherghetta,
Laurent Lellouch, Joe Lykken and Riccardo Rattazzi for comments and 
conversations. 
SFK would like to thank CERN and Fermilab for hospitality extended.

\end{document}